# Configured Grant Scheduling for the Support of TSN Traffic in 5G and Beyond Industrial Networks

M.C. Lucas-Estañ[1], A. Larrañaga[2], J. Gozalvez[1], I. Martínez[2]
[1]UWICORE Laboratory, Universidad Miguel Hernández de Elche (UMH), Elche 03202, Spain
[2]HW and Communication Systems Area, Ikerlan Technology Research Centre, Mondragón 20500, Spain
{m.lucas, j.gozalvez}@umh.es, {ana.larranaga, imartinez}@ikerlan.es

***Abstract*— 5G and beyond networks can facilitate the digital transformation of manufacturing and support more flexible and reconfigurable factories with ubiquitous mobile connectivity. This requires integrating 5G networks with industrial networks that increasingly rely on TSN (Time Sensitive Networking) to support deterministic communications with bounded latencies. Deterministic communications are critical for many industrial applications, but 5G does not natively support deterministic communications. To address this limitation, this study proposes the coordination of the 5G and TSN schedulers and presents a novel 5G configured grant scheduling scheme to support TSN traffic. The scheme uses information about the characteristics of the TSN traffic (packet size, periodicity, and arrival time) to coordinate its scheduling decisions with the TSN scheduler. The study demonstrates that the proposed scheme outperforms the state-of-the-art in the capacity to support multiple TSN traffic flows with different periodicities.**

***Keywords—5G, TSN, industrial networks, 5G-TSN integration, scheduling, deterministic communications, Configured Grant.***

## I. Introduction

The digital transformation of manufacturing requires flexible and reconfigurable factories that efficiently integrate (mobile) cooperative robots and cyber-physical systems. Industrial wired networks increasingly rely on Time Sensitive Networking (TSN) to support mixed traffic flows and deterministic communications with bounded latencies. 5G does not natively support deterministic communications but provides the mobility support and reconfigurability required by future factories. In this context, 5G Alliance for Connected Industries and Automation (5G-ACIA) identifies the integration of 5G and beyond with TSN as a fundamental step to support the digital and data-centric transformation of manufacturing [1].

The 3GPP defines the framework for integrating 5G and TSN where the 5G system operates as a logical bridge in a TSN network. However, 3GPP does not specify how the 5G and TSN schedulers should be integrated, and this is critical to satisfy the end-to-end (E2E) latency requirements of industrial TSN traffic over 5G, especially under mixed traffic flows. A first proposal is presented in [2] where authors introduce a scheduler that preempts enhanced mobile broadband (eMBB) traffic when prioritized deterministic traffic must be transmitted. The scheduler seeks minimizing the impact of preemption on the eMBB throughput while guaranteeing the latency requirements of deterministic traffic. The proposal in [3] combines semi-static and dynamic scheduling to support TSN traffic over 5G. The proposal pre-allocates periodic radio resources for the transmission of TSN traffic. If the TSN traffic varies and the pre-allocated resources are insufficient to support it, additional resources are dynamically scheduled to support the additional TSN traffic. A similar approach is followed in [4] to support mixed traffic flows (periodic time-critical, event-triggered time-critical and best-effort). The proposal pre-allocates resources for the periodic time-critical traffic, reserves resources in each slot for potential event-triggered time-critical traffic, and dynamically schedules best-effort traffic. Best-effort traffic can also be scheduled over unutilized reserved resources with preemption applied if new event-triggered traffic is generated.

The proposals in [3] and [4] advocate for the use of semi-static scheduling (Configured Grant -CG- for uplink traffic or Semi-Persistent Scheduling -SPS- for downlink) to support periodic deterministic TSN traffic. The proposals assign one configured uplink (UL) grant to each TSN traffic flow. Each grant periodically pre-allocates radio resources to each TSN flow with the same periodicity as the TSN traffic. Existing proposals manage each grant or TSN flow independently of each other, and this can generate scheduling conflicts if each TSN flow generates traffic with different periodicities. Fig. 1 illustrates such conflicts that occur when two or more TSN traffic flows are pre-allocated the same resources at a certain time. To address this challenge, this paper proposes a novel 5G configured grant scheduling scheme that can efficiently manage multiple UL TSN flows with different periodicities (it is important to note that periodic deterministic traffic is the most common traffic generated in industrial environments [5]). The proposed scheme uses information about each TSN flow (periodicity, packet size and packet arrival time) to coordinate its scheduling decisions with the TSN network and avoid scheduling conflicts. The scheme avoids conflicts between TSN flows with different periodicities by assigning several configured UL grants to each TSN flow. The scheme identifies a common period (hyperperiod or HP) for all TSN traffic flows, and configures a different UL grant for each packet of a TSN flow in the *HP*. Each configured UL grant allocates the radio resources that minimize the latency of the corresponding packet, and this radio resource allocation repeats periodically with a periodicity *HP*. The conducted evaluation shows that the proposed CG scheduling scheme avoids scheduling conflicts, and considerably increases the number of TSN traffic flows that 5G can satisfactorily support, i.e., for which their E2E latency requirement is met.

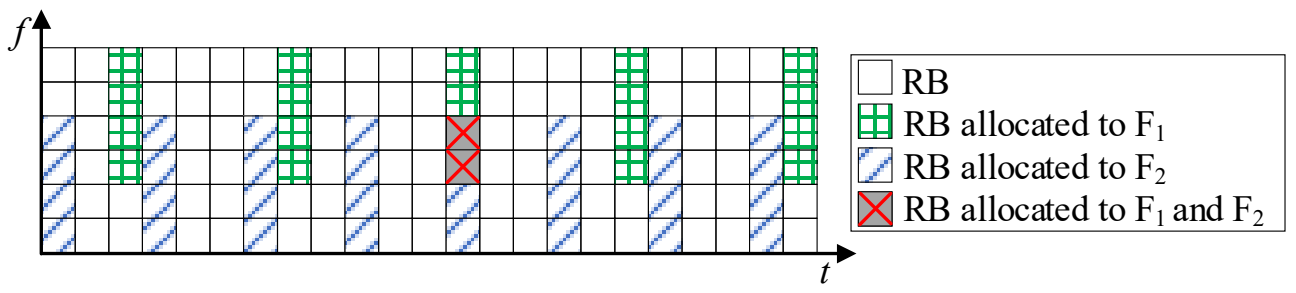

Fig. 1. Scheduling conflict between two TSN flows with different periodicities.

## II. 5G-TSN INTEGRATION MODEL

A TSN network is composed of end devices and bridges interconnected using standard Ethernet links. The TSN bridges are Ethernet bridges with special features to guarantee deterministic communications. These features include (among others) a strict time synchronization using the IEEE 802.1AS standard, and a priority-based scheduler that reserves specific transmission intervals for high priority traffic using the IEEE 802.1Qbv standard.

The 3GPP defines in [6] the framework for the integration of 5G and TSN. 3GPP defines that a 5G network should integrate in a TSN network as a logical bridge. The integration model is depicted in Fig. 2, and the 5G System (5GS) is referred to as 5GS Bridge. The 3GPP integration model includes TSN translators (TT) at the interconnection points between the 5G and TSN networks, i.e. at the User Equipment or UE (the device-side TT or DS-TT) and the 5G Core Network (the network-side TT or NW-TT). The TTs act as ingress and egress ports of the logical bridge, and their main functions are to understand, translate and execute TSN configuration messages received from the TSN network. The Central Network Configuration (CNC) in a TSN network is in charge of the centralized management of the integrated 5G-TSN network. The CNC is aware of the communication requirements of the end-devices, the status and capabilities of all (TSN and 5GS) bridges in the network, and the delay suffered at each bridge and link of the integrated 5G-TSN network. The CNC uses this information to schedule transmissions following the IEEE 802.1Q standard. In particular, it establishes the communication path and the time instants at which each packet of a TSN flow should arrive and depart from each TSN and 5GS bridge to ensure that the end-to-end requirements of the TSN traffic are met. The CNC sends the scheduling decisions to the TSN and 5GS bridges in the network for their configuration.

It is important to highlight that 3GPP or TSN standards do not specify a scheduling mechanism for the TSN traffic over a 5G network. Such scheduling should be done so that the arrival and departure times at the 5GS bridge established by the CNC are met. To this end, the CNC provides information to the 5G network regarding the packet arrival time, the flow direction, the survival time [1], and the periodicity of the packets[2] for each TSN flow. The packet arrival time refers to the time when the first packet of the flow arrives at the 5GS bridge (specifically, at the ingress port of the 5GS bridge that is the UE in UL transmissions). In addition, a 5G QoS profile is assigned to each TSN flow. The 5G QoS profile includes the flow's priority level, the maximum data burst volume (MDBV), and its required packet error rate (PER) and packet delay budget (PDB). The PDB indicates the maximum latency for a packet to be transmitted over the 5G network.

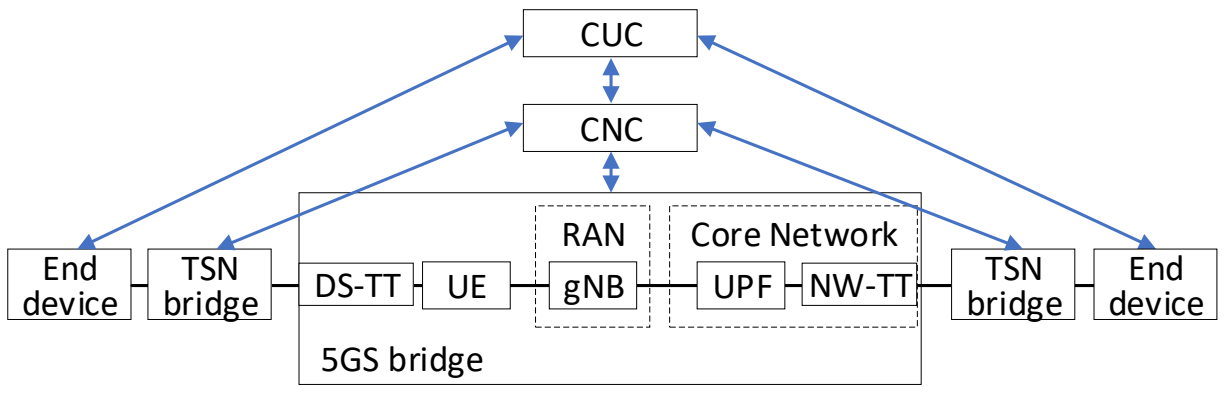

Fig. 2. 5G-TSN integration framework proposed by the 3GPP [6].

## III. 5G NR SCHEDULING FOR TSN TRAFFIC

This section presents a novel 5G Configured Grant scheduling scheme capable to support multiple periodic TSN flows in the integrated 5G-TSN framework depicted in Fig. 2. We refer to the proposed scheme as Optimum Flexible configured grAnt Scheduling for TSN traffic (O-FAST). The proposed scheme pre-allocates radio resources periodically to the TSN flows and configures multiple UL grants for each TSN flow if required. The scheme avoids scheduling conflicts between TSN flows with different periodicities (see Fig. 1). To avoid the conflicts, the proposed scheme exploits the information provided by the CNC to satisfy the latency requirements of all TSN flows. In particular, the scheme utilizes the information about the arrival time of the first packet of each TSN flow at the 5GS bridge, and the size and periodicity of the packets of each TSN flow.

We consider a 5G network that must support $N_F$ TSN traffic flows. Each TSN flow $F_i$ (with $i$=1,..,$N_F$) is characterized by the transmission of periodic packets with size $size_i$ and periodicity $p_i$. We denote as $pkt_{i,j}$ the packets of a TSN flow $F_i$, where $i$ and $j$ indicate that it is the $j^{th}$ packet in the TSN flow $F_i$, and $j$=1,2,…. The 5G network knows the arrival time of the first packet $pkt_{i,1}$ of a TSN flow $F_i$ at the 5GS bridge, which is denoted with $A_{i,1}$, and the maximum latency that must be guaranteed in the 5G network for the packets of a TSN flow $F_i$, referred to as $l_i^{5G}$. Based on the previous information, the arrival time for any packet $pkt_{i,j}$ of a TSN flow $F_i$ can be calculated as $A_{i,j} = A_{i,1} + (j\text{-}1)\cdot p_i$.

O-FAST first computes the hyperperiod ($HP$) that is defined as the least common multiple of the periodicities of all the TSN flows: $HP$=LCM$\{p_i\}$, $\forall i$ = 1,..,$N_F$. O-FAST then schedules the packets of all the TSN flows within $HP$, and repeats the scheduling in following $HPs$ since all the TSN flows repeat their traffic pattern with periodicity $HP$. For a TSN flow $F_i$, O-FAST allocates the radio resources to each packet $pkt_{i,j}$ in a TSN flow $F_i$ that minimize the latency considering the available radio resources. This is not the case with commonly used CG schemes that allocate resources with periodicity $p_i$ for the transmission of all the packets of a TSN flow $F_i$. O-FAST selects the radio resources reserved to each packet $pkt_{i,j}$ of a TSN flow $F_i$ to avoid scheduling conflicts among different flows (Fig. 1). To this aim, O-FAST divides the scheduling problem into several sub-problems with lower computational cost. It creates different groups $G_z$ of packets from all TSN flows $F_i$ included in an $HP$. Each group includes the packets $pkt_{i,j}$ (with $i$=1,..,$N_F$ and $j$=1,2,…) whose transmission could overlap in time considering their arrival time $A_{i,j}$ at the 5GS bridge and the maximum latency $l_i^{5G}$ that must be guaranteed. The transmission of two packets $pkt_{i,j}$ and $pkt_{m,n}$ can overlap in time when $A_{i,j} \leq A_{m,n} \leq A_{i,j} + l_i^{5G}$ or $A_{i,j} \leq A_{m,n} + l_m^{5G} \leq A_{i,j} + l_i^{5G}$. O-FAST then defines an optimization problem for each group $G_z$ to identify the radio resources that must be allocated to each packet $pkt_{i,j}$ (with $i$=1,..,$N_F$ and $j$=1,2,….) within $G_z$ to avoid scheduling conflicts.

O-FAST needs to identify the radio resources that must be allocated for the transmission of each packet $pkt_{i,j}$ within $G_z$ to guarantee the maximum latency $l_i^{5G}$ established by the TSN scheduler. The number of radio resources needed to transmit a packet $pkt_{i,j}$ of size $size_i$ is calculated as:

[1] The survival time indicates the time period an application can survive without receiving any data packet.

[2] This information is included in the Time Sensitive Communication Assistance Information (TSCAI) message [6].

$$d_i = \left\lceil \frac{[tbs_i(size_i+header)+CRC]\cdot 8}{R\cdot Q_m\cdot N_{sc,RB}} \right\rceil \quad (1)$$

where *header* is the length of the IPv4 header (in bits) and *CRC* is the length of the cyclic redundancy check (CRC) code in bytes. $tbs_i(size_i+header)$ is the smallest transport block size that can transmit a packet of size $size_i+header$. $R$ and $Q_m$ are the coding rate and modulation order, respectively, used to transmit a packet. $N_{sc,RB}$ is the number of subcarriers in a resource block or RB of the 5G NR grid. 5G NR defines a grid organized in RBs in the frequency domain and OFDM symbols in the time domain. An RB is composed of 12 subcarriers in the frequency domain, and the number of RBs depends on the bandwidth and is represented by $R_{BW}$. A radio resource consists of an RB in the frequency domain and an OFDM symbol in the time domain. We number the symbols (starting from 0) within an *HP*, and each symbol has a time duration of $t_{sym}$ that depends on the 5G NR numerology. OFDM symbols are organized in slots that consist of 14 or 12 symbols depending on the 5G NR numerology.

O-FAST establishes that when the number of radio resources $d_i$ needed for the transmission of a packet $pkt_{i,j}$ is lower than $R_{BW}$, the packet $pkt_{i,j}$ will receive $d_i$ RBs in the same symbol. When $d_i$ is higher than $R_{BW}$, the packet $pkt_{i,j}$ will receive $R_{BW}$ in $\lceil d_i/R_{BW} \rceil$ consecutive symbols. The number of symbols and RBs needed to satisfy $d_i$ for a packet $pkt_{i,j}$ is represented by $d_i^S$ and $d_i^R$, respectively. O-FAST allocates to each packet $pkt_{i,j}$ the radio resources that minimize its latency considering the arrival time at the 5GS bridge $A_{i,j}$ and the number of radio resources $d_i$ needed to transmit each packet $pkt_{i,j}$ in a group $G_z$. We denote with $l_{i,j}$ the latency experienced by the packet $pkt_{i,j}$ in the 5G network. O-FAST then minimizes the sum of the latency $l_{i,j}$ experienced by all the packets $pkt_{i,j}$ in $G_z$, with $i \in [1, N_F]$ and $j \in [1, N_{pkt}^{i,z}]$, where $N_{pkt}^{i,z}$ represents the number of packets for the TSN flow $F_i$ within $G_z$:

$$\min \sum_{i=1}^{N_F} \sum_{j=1}^{N_{pkt}^{i,z}} l_{i,j} \quad (2)$$

The latency experienced by a packet in a 5G network can be calculated as the sum of the latency experienced in the radio, transport, and core networks. This study considers a 5G Non-Public Network (NPN) with a core network implemented in the factory premises. In this scenario, the latency experienced in the transport and core network can be considered negligible compared to the latency experienced in the radio network [7]. $l_{i,j}$ can then be estimated as the difference between the time the packet is received in the gNB and the arrival time $A_{i,j}$ of the packet at the 5GS bridge (it is important to note that the ingress port is the UE for UL transmissions). The time at which the packet is received in the gNB depends on: 1) the time at which the transmission of packet $pkt_{i,j}$ starts, 2) the transmission time of the packet $pkt_{i,j}$ through the radio channel that is given by $d_i^S\cdot t_{sym}$, and 3) the processing time $t_{gNB,rx}$ needed to decode the packet at the receiver. If we represent as $s_{i,j}$ the first symbol allocated to a packet $pkt_{i,j}$, the time at which the transmission of packet $pkt_{i,j}$ starts is given by $s_{i,j}\cdot t_{sym}$. Then, $l_{i,j}$ can be expressed as:

$$l_{i,j} = s_{i,j}\cdot t_{sym} + d_i^S\cdot t_{sym} + t_{gNB,rx} - A_{i,j} \quad (3)$$

$s_{i,j}$ in (3) can be expressed as:

$$s_{i,j} = \sum_{s=0}^{S} s\cdot X_s^{i,j} \quad (4)$$

with $X_s^{i,j}$ a binary variable equal to 1 when $s$ is the first symbol allocated for the transmission of packet $pkt_{i,j}$ and 0 in other case. $S$ in (4) represents the length of the *HP* expressed in number of symbols. Using (3) and (4), the objective function of O-FAST defined in (2) can now be expressed as:

$$\min \sum_{i=1}^{N_F} \sum_{j=1}^{N_{pkt}^{i,z}} \left( \sum_{s=0}^{S} s\cdot X_s^{i,j} \right)\cdot t_{sym} + d_i^S\cdot t_{sym} + t_{gNB,rx} - A_{i,j} \quad (5)$$

with $X_s^{i,j} \in \{0,1\}$, $\forall i \in \{1,\ldots, N_F\}$, $\forall j \in \{1,\ldots, N_{pkt}^{i,z}\}$, and $\forall s \in \{1,\ldots, S\}$.

The solution to the optimization problem must satisfy that the latency $l_{i,j}$ experienced by a packet $pkt_{i,j}$ in the 5GS is equal to or lower than the maximum latency established by the CNC of the TSN network ($l_i^{5G}$). This constraint to the optimization problem is expressed as:

$$\left( \sum_{s=0}^{S} s\cdot X_s^{i,j} \right)\cdot t_{sym} + d_i^S\cdot t_{sym} + t_{gNB,rx} - A_{i,j} \le l_i^{5G}, \quad \forall j \in \{1,\ldots, N_{pkt}^{i,z}\}, \forall i \in \{1,\ldots, N_F\}. \quad (6)$$

O-FAST takes into account the processing time $t_{UE,tx}$ required by the transmitter (the UE) to generate and encode the packet. The radio resources allocated to the transmission of a packet $pkt_{i,j}$ must then be in symbols after $A_{i,j}+t_{UE,tx}$, and this constraint is expressed as:

$$X_s^{i,j}=0, \quad \forall s < \left\lceil \frac{A_{i,j}+t_{UE,tx}}{t_{sym}} \right\rceil, \quad \forall j \in \{1,\ldots, N_{pkt}^{i,z}\}, \forall i \in \{1,\ldots, N_F\} \quad (7)$$

5G NR can reserve some symbols within a slot for the transmission of control channels. O-FAST considers that the first $sym_{DL\text{-}Ctrl}$ symbols of each slot are reserved for control channels in DL and the last $sym_{UL\text{-}Ctrl}$ symbols of each slot are reserved for control channels in UL. To avoid allocating symbols reserved for DL control channels, we define the following constraint for O-FAST:

$$X_s^{i,j}=0, \quad \forall s \mid 0 < \mathrm{mod}(s, 14) \le sym_{DL\text{-}Ctrl}, \quad \forall j \in \{1,\ldots, N_{pkt}^{i,z}\}, \forall i \in \{1,\ldots, N_F\} \quad (8)$$

The 3GPP standards establish that all the radio resources allocated for the transmission of a packet must be in the same slot. Assuming a 5G NR numerology with a normal cyclic prefix, each slot consists of 14 symbols. $\lfloor s/14 \rfloor$ is the number of complete slots from the beginning of the *HP* to the symbol $s$ that represents the first symbol allocated for the transmission of the packet. $s - \lfloor s/14 \rfloor\cdot 14$ indicates the number of the symbol $s$ within the current slot. To ensure that the $d_i^S$ symbols allocated for the transmission of packet $pkt_{i,j}$ are within the same slot and do not include the $sym_{UL\text{-}Ctrl}$ symbols reserved for UL control channels, the solution to the optimization problem defined for O-FAST must satisfy:

$$\sum_{s=0}^{S} (s - \lfloor s/14 \rfloor\cdot 14)\cdot X_s^{i,j} \le 14 - sym_{UL\text{-}Ctrl} - d_i^S, \quad \forall j \in \{1,\ldots, N_{pkt}^{i,z}\}, \forall i \in \{1,\ldots, N_F\} \quad (9)$$

O-FAST aims to successfully serve all the packets in $G_z$. As a result, one variable $X_s^{i,j}$ with $s \in \{1,\ldots,S\}$ must be equal to one for all packets $pkt_{i,j}$ in $G_z$, which is expressed as:

$$\sum_{s=0}^{S} X_s^{i,j} = 1, \quad \forall j \in \{1,\ldots,N_{pkt}^{i,z}\}, \forall i \in \{1,\ldots,N_F\} \tag{10}$$

In order to avoid scheduling conflicts, O-FAST must ensure that the same resources are not allocated for the transmission of more than one packet. To this end, O-FAST does not allocate more than $R_{BW}$ RBs in each symbol, which is expressed in (11). If this constraint is fulfilled, it will be possible to allocate different RBs for the transmissions of different packets, and finally avoid conflicts:

$$\sum_{i=1}^{N_F} \sum_{j=1}^{N_{pkt}^{i,z}} X_s^{i,j} \cdot d_i^R \le R_{BW}, \forall s \in \{1,\ldots,S\} \tag{11}$$

If a packet demands more than $R_{BW}$ resources, the $d_i$ resources must be allocated in $d_i^S$ consecutive symbols, which is expressed with the following constraint:

$$X_s^{i,j} \cdot R_{BW} + \sum_{s'=s+1}^{s+d_i^S-1} \sum_{\substack{m\in[1,N_F] \\ m\neq i}} \sum_{n\in[1,N_{pkt}^{i,z}]} X_{s'}^{m,n} \le R_{BW}, \tag{12}$$

$\forall s \in \{1,\ldots,S\}, \forall j \in \{1,\ldots,N_{pkt}^{i,z}\}, \forall i \in \{1,\ldots,N_F\}$ with $d_i^S > 1$

To summarize, the O-FAST scheduling proposal is defined with the objective function in (5), and the constraints in (6), (7), (8), (9), (10), (11), and (12). The optimization problem results in a Binary Integer Programming (BIP) problem where the unknown variables are binary variables.

## IV. Evaluation Scenario

We evaluate the proposed 5G-TSN scheduling scheme in an industrial plant scenario covered by a single-cell 5G private network (NPN) that is integrated with an industrial TSN network following the architecture depicted in Fig. 3. The scenario considers the deployment of a closed-loop supervisory application [8] where a PLC (Programmable Logic Controller) receives monitoring data from $N_F$ sensors ($S_1, S_2, \ldots, S_{NF}$), and transmits a command to actuator $A$ in Fig. 3. $N_F$ TSN flows from the sensors to the PLC pass through the 5GS bridge. We evaluate the performance of the scheduling scheme with values of $N_F$ ranging between 10 and 30. For each TSN flow $F_i$, the corresponding sensor generates packets of size $size_i$ with a periodicity $p_i$. Following [8], $size_i$ and $p_i$ for each $F_i$ are randomly selected between 40 and 250 bytes, and between 4 and 20 ms, respectively. Each packet must be received before the next packet is generated [8]. Consequently, the E2E latency requirement ($l_i^{e2e}$) for each TSN flow $F_i$ is set equal to $p_i$. The time at which a TSN packet $pkt_{i,j}$ arrives to the 5GS ($A_{i,j}$) and the maximum latency that 5GS must guarantee ($l_i^{5G}$) are computed using (13) and (14), respectively, derived following the analysis in [9]:

$$A_{ij} = l_{sensor} + l_{link\text{-}ingress_i} + l_{TSN\text{-}ingress_i} + t_{DS\text{-}TT} \tag{13}$$

$$l_i^{5G} \le l_i^{e2e} - A_{ij} - l_{link\text{-}egress_i} - l_{TSN\text{-}egress_i} - t_{NW\text{-}TT} - l_{PLC} \tag{14}$$

In (13) and (14), $l_{sensor}$ and $l_{PLC}$ represent the application processing time at the sensor and PLC, respectively, $l_{TSN\text{-}ingress_i}$ and $l_{TSN\text{-}egress_i}$ represent the latency experienced by the TSN flow $F_i$ in the TSN bridges in the path between the sensor $i$ and the 5GS bridge, and between the 5GS bridge and the PLC, respectively, and $l_{link\text{-}ingress_i}$ and $l_{link\text{-}egress_i}$ represent the propagation time it takes a packet to travel through the links between the sensor $i$ and the 5GS bridge, and between the 5GS bridge and the PLC, respectively. $t_{DS\text{-}TT}$ and $t_{NW\text{-}TT}$ are the processing times at the DS-TT and NW-TT.

We simulate a 5G network in Matlab with 20 MHz, a sub-carrier spacing (SCS) equal to 30 kHz, and operating in TDD mode following recommendations in [10] and [11]. The UEs (industrial devices) are deployed under Line of Sight (LoS) conditions with the gNB, and utilize MCS12 (Modulation and Coding Scheme) from Table 1 in [12] for their transmissions. This MCS guarantees a good trade-off between robustness and transmission rate under LoS conditions as it uses 16QAM (with modulation order $Q_m$=4) and coding rate $R$=434/1024.

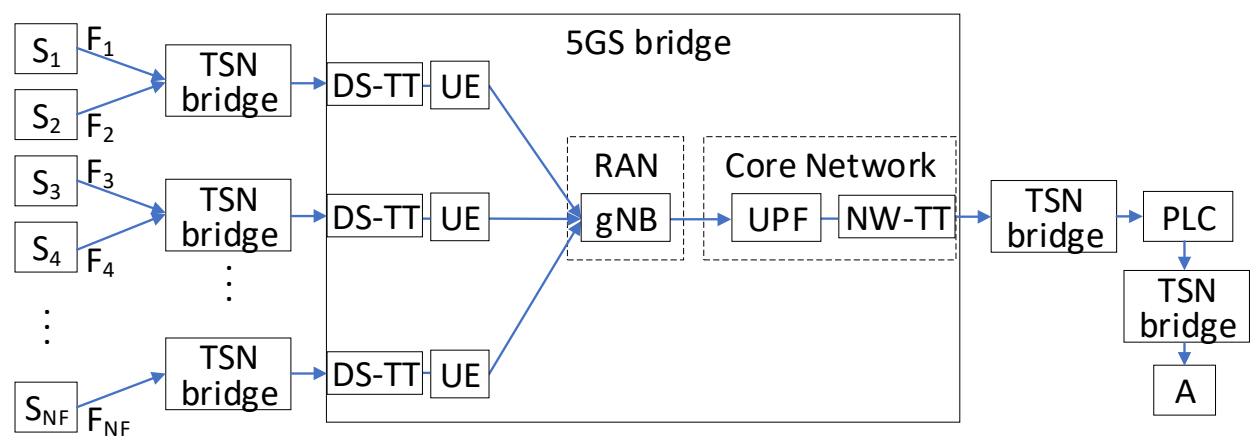

Fig. 3. 5G-TSN integrated network under evaluation.

## V. Performance Results

We compare the performance of the proposed O-FAST scheduling scheme with a commonly used CG (Configured Grant) scheduling scheme [3][4]. The reference scheme configures a single UL grant for each TSN flow $F_i$, and periodically assigns radio resources to each flow. The periodicity is set equal to the periodicity $p_i$ of each TSN flow $F_i$. The scheme assigns each TSN flow the number of radio resources necessary to satisfy its demand $d_i$. For a fair comparison with O-FAST, the reference scheme serves first those TSN flows with more stringent latency requirements.

Fig. 4 depicts the percentage of scheduling problem for which O-FAST and the reference scheme achieve a solution that meets the latency requirement for all TSN flows as a function of the number $N_F$ of TSN flows. Fig. 4 shows that O-FAST outperforms the reference scheme. The reference scheme significantly degrades its capacity to satisfy the latency requirements of all the TSN flows as the number $N_F$ of TSN flows increases. This is due to the conflicts that can occur when a CG scheduling scheme periodically assigns radio resources to each flow independently of each other, and each flow has different periodicity. In this case, two or more flows may receive the same radio resources at the same time (Fig. 1), and this results in packet collisions as shown in Fig. 5. The figure shows that the percentage of packets that collide with the reference scheme increases with the number $N_F$ of TSN flows. Such collisions are critical for industrial applications that can demand reliability levels as high as $10^{-9}$

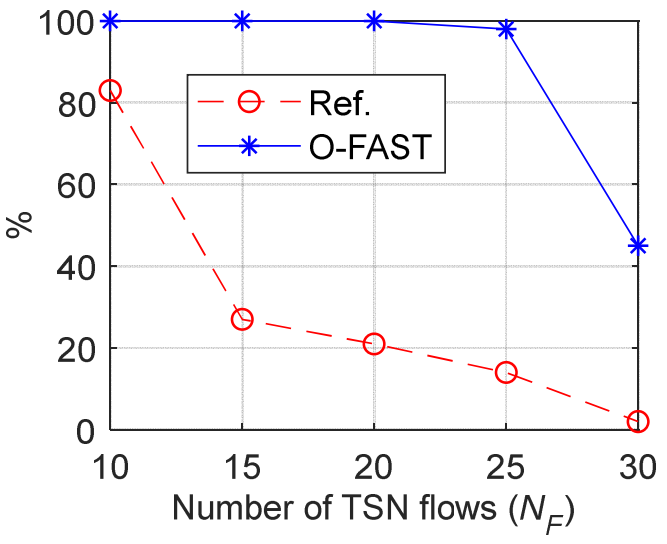

Fig. 4. Percentage of scheduling problems for which the latency requirements for all the TSN flows are satisfied as a function of $N_F$.

[8]. On the other hand, O-FAST guarantees the absence of packet collisions as its resource allocation avoids conflicts between TSN flows independently of the number of flows and their periodicity. O-FAST can then satisfy the latency requirements of all TSN flows for up to 25 TSN flows. The capacity of O-FAST to satisfy the latency requirements of all TSN flows decreases when $N_F$ increases to 30. This is not due to packet collisions between different TSN flows, but to the fact that there are no feasible solutions that can satisfy the stringent latency requirements for all TSN flows when $N_F$ increases to 30. In any case, O-FAST still significantly outperforms the reference scheme for $N_F$=30.

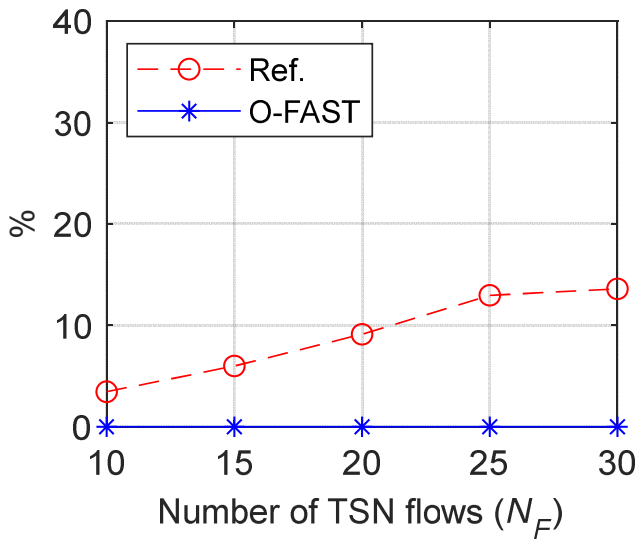

Fig. 5. Percentage of packet collisions as a function of $N_F$.

Fig. 6 compares the average latency experienced in the 5GS by the packets of the different TSN flows with O-FAST and the reference scheme. The figure shows that O-FAST significantly reduces the average latency compared to the reference scheme. This is because the reference scheme allocates the radio resources that minimize the latency for the first packet of each TSN flow. The allocated radio resources then repeat with a periodicity equal to the periodicity $p_i$ of the TSN flow, and do not change even if there are other radio resources available that could decrease the latency for following packets of the TSN flow. This results in that all packets of a TSN flow experience the same latency with the reference scheme, i.e., the jitter is null. On the other hand, O-FAST flexibly allocates to each packet the radio resources that minimize the latency based on the current resource availability. The flexibility of O-FAST comes at the expense of a small jitter (55-300 µs when $N_F$ varies between 10 and 30 flows). We should note that the jitter can be eliminated at the NW-TT as it can hold the packets and forward them to the next node at the departure time ($A_{i,j}+l_i^{5G}$) established by TSN.

Previous results have shown that O-FAST can avoid scheduling conflicts (and hence packet collisions) and significantly reduce the latency compared to traditional 5G CG scheduling schemes. This results in that O-FAST can effectively support large number of TSN flows with different periodicities. These gains come at the expense of a larger computational cost to decide the resource allocations. The reference scheme can take its scheduling decisions at the millisecond level. On the other hand, O-FAST can require between 1 minute and 1.5 hours to solve the scheduling problem in scenarios with $N_F$ between 10 and 30 flows[3]. These values can be reduced with sub-optimal scheduling solutions that are left for future work. However, it is important to highlight that O-FAST does not need to be executed in real time since the scheduling solutions can be derived during the offline planning phase of the industrial plant just like it is done for the scheduling of TSN networks. Such planning is done when the industrial plant is deployed, and the scheduling decisions are maintained as long as the industrial layout is maintained. As a result, O-FAST can compute offline its scheduling decisions, and these decisions are maintained for long periods of time.

## VI. Conclusions

This study has presented a novel 5G configured grant scheduling scheme to support TSN traffic characteristic of industrial networks. The scheme coordinates its scheduling decisions with the TSN network and uses information about each TSN traffic flow to schedule multiple TSN flows with different periodicities. The proposed scheme assigns multiple configured UL grants to each TSN flow and adapts the resource allocation on an hyperperiod basis to avoid scheduling conflicts among TSN flows. The study shows that, compared to existing 5G CG scheduling schemes, the proposed scheme can significantly augment the number of TSN flows that satisfy their E2E latency requirements when transmitted over 5G. The proposed scheme also significantly reduces the average 5G E2E latency at the expense of larger computational times to decide the resource allocations. We should though note that the scheduling decisions are planned offline (non-real-time) when the industrial layout is done, just like it is done with industrial TSN networks.

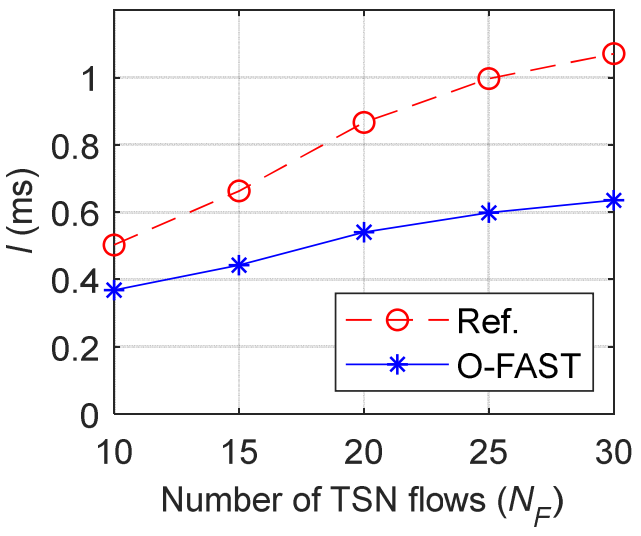

Fig. 6. Average latency ($l_i^{5G}$) experienced by the packets of the different TSN flows as a function of $N_F$.

## Acknowledgment

This work has been funded by European Union's Horizon Europe Research and Innovation programme under the Zero-SWARM project (No 101057083), by MCIN/AEI/ 10.13039/501100011033 through the project PID2020-115576RB-I00, and by Generalitat Valenciana (CIGE/2022 /17), and UMH's Vicerrectorado de Investigación grants.

[3] The scheduling problems are solved using Matlab's toolbox Parallel Computing, Optimization y Symbolic Math. The values reported have been measured with the toolbox executed in a server with an Intel(R) Core (TM) i7-5930K @ 3.50GHz CPU and 16GB RAM.